\documentclass[oldversion,letter]{aa}
\usepackage{graphicx}
\usepackage{txfonts}

\usepackage{natbib}
\bibpunct{(}{)}{;}{a}{}{,} 

\begin{document}

\title{X-ray accretion signatures in the close CTTS binary V4046 Sgr}
\author{H.~M. G\"{u}nther\inst{1} \and C. Liefke\inst{1} \and J.~H.~M.~M. Schmitt\inst{1} \and J. Robrade\inst{1} \and J.-U. Ness\inst{2}}
\institute{
Hamburger Sternwarte,
Gojenbergsweg 112,
D-21029 Hamburg, Germany
\and
Arizona State University,
School of Earth and Space Exploration,
Tempe, AZ, 85287-1404, USA
}

\mail{H.M. G\"{u}nther, \\moritz.guenther@hs.uni-hamburg.de}

\date{Received 28 August 2006/ Accepted 26 September 2006}

\abstract{We present {\it Chandra} HETGS observations of the classical T Tauri star (CTTS) \protect{\object{V4046~Sgr}}. 
The He-like triplets of \ion{O}{vii}, \ion{Ne}{ix}, and \ion{Si}{xiii} are clearly detected. Similar to the CTTS
TW~Hya and BP~Tau, the forbidden lines of
\ion{O}{vii} and \ion{Ne}{ix} are weak compared to the intercombination line, indicating high plasma densities in the X-ray emitting regions. 
The \ion{Si}{xiii} triplet, however, is within the low-density limit, in agreement with the predictions of the accretion funnel infall 
model with an additional stellar corona. V4046~Sgr is the
first close binary exhibiting these features. Together with previous high-resolution
X-ray data on \protect{\object{TW~Hya}} and \protect{\object{BP~Tau}}, and in contrast to \protect{\object{T~Tau}}, now three out of four CTTS show evidence of accretion funnels.}

\maketitle

\section{Introduction}

T Tauri stars are young (age $<10$~Myr), low-mass ($M_*<3M_{\sun}$) stars evolving
down to the main sequence in the Hertzprung-Russell diagram;
a detailed review of pre-main sequence stellar evolution is given by \citet{1999ARA&A..37..363F}. 
The CTTS subclass shows signs of active accretion from a circumstellar disk as proven by strong infrared excess and 
a large H$\alpha$ equivalent 
width ($>$10~\AA{}). Current models of CTTS assume a dipolar stellar magnetic field, disrupting the disk near the corotation 
radius, where ionised material is loaded onto magnetic field lines and is accelerated to essentially free-fall velocity 
along a magnetic funnel \citep{1994ApJ...429..781S}. A shock is formed close to the stellar surface and the material passing 
through the shock is heated up to temperatures in the MK range.  The accretion
geometries for binaries are understood less well theoretically, but see \citet{1996ApJ...467L..77A}.

The X-ray emission from TTS was first detected with the {\em Einstein} Observatory, and extensive X-ray studies
of a variety of star forming regions were undertaken with {\em ROSAT}. However, to what extent the observed X-ray emission can actually be attributed to accretion processes remained unclear and is the subject of some debate, since  a 
central result of stellar X-ray astronomy states that all ``cool stars'', i.e., stars with outer convective envelopes, are 
surrounded by hot coronae \citep{2004A&A...417..651S}. The usual interpretation of this finding is that turbulence in the outer
convective zone leads to the production of magnetic fields and the ensuing activity, making it natural to 
attribute the observed X-ray emission from CTTS to the phenomena of magnetic activity rather than of accretion.

Imaging studies with low/medium spectral resolution allow large samples of TTS to be studied simultaneously as was recently performed for the Orion nebula by the COUP project \citep{2005sfet.confE..39F}.  In such studies, stars with and without accretion disks are difficult to distinguish; however, statistical studies seem to indicate some differences between CTTS and those TTS without accreting disks \citep{2000A&A...356..949S, 2001A&A...377..538S, 2003A&A...402..277F}.

Only with the high spectral resolution of grating spectra, as available for a few individual CTTS observed with 
\emph{Chandra} and \emph{XMM-Newton}, is it possible to identify spectroscopic signatures taht distinguish accretion and coronal emission and to reliably derive the physical properties of the emission region through temperature, density, and abundance diagnostics.  Specifically, a high spectral resolution survey of 48 stellar coronae by \citet{2004A&A...427..667N} showed that most of the coronal densities are compatible with the low-density limit within the error ranges.

Only two clear exceptions among cool stars are known to date, the CTTS TW~Hya and BP~Tau. 
In the \emph{Chandra} HETGS spectrum of TW~Hya, \citet{2002ApJ...567..434K} found unusually high densities and interpreted them as the signature of an accretion funnel; this was then confirmed by an \emph{XMM-Newton} observation described by \citet{twhya}. The same low f/i-ratios for the \ion{O}{vii} 
and \ion{Ne}{ix} triplets were observed in \emph{XMM-Newton} RGS spectra of the CTTS BP~Tau \citep{bptau}, while the \emph{XMM-Newton} EPIC spectra also clearly
show the presence of rather hot plasma.  The authors interpret the BP~Tau data
in terms of accretion with an additional 
active corona; similar conclusions were drawn by \citet{Rob0507} in a comparative study of the high and medium resolution X-ray data of four CTTS.
The interpretation of the f/i-ratios in terms of density is not unambiguous.
\citet{Ness0510} presented a careful examination of density-sensitive iron lines 
in TW~Hya and also showed that neither the UV flux nor line scattering affects the He-like triplets. 

In this letter we present {\it Chandra} HETG
spectra of the CTTS V4046~Sgr, which we show to closely resemble the X-ray spectra of TW~Hya and  BP~Tau and to be the third star with very low  He-like f/i-ratios.  We therefore suggest that accretion-related X-ray emission may in fact be quite common in CTTS.

\section{V4046~Sgr: Observations and data reduction}

V4046~Sgr is a nearby CTTS at a distance between
42 pc \citep{1990A&A...234..230H} and 83 pc \citep{2000IAUS..200P..28Q}. 
The star is isolated from any dark or molecular cloud and has negligible extinction. 
The CTTS nature of V4046~Sgr has been established through its
variable H$\alpha$ equivalent width \citep[30--120~\AA,][]{1986IrAJ...17..294B} and a strong infrared excess ,IRAS-detected \citep{1986niia.conf..107D}, while other IR observations are also consistent with a disk surrounding V4046~Sgr \citep{1990A&A...234..230H}. 
V4046~Sgr itself is a spectroscopic binary with a well-determined orbital period of 2.4213459 days \citep{2004A&A...421.1159S}, consisting of a K7\,V and a K5\,V 
star with masses $M_1=0.86M_{\sun{}}$ and $M_2=0.69M_{\sun{}}$ at an orbital separation of less than 10~$R_{\sun}$ \citep{2000IAUS..200P..28Q}. The observed 
spectral energy IR distribution is consistent with an inner disk radius of 1.8~AU \citep{1997AJ....114..301J}, the disk surrounding
V4046~Sgr is thus circumbinary \citep{2000IAUS..200P..28Q}.
The viewing angle of the system is $i=35$--$45^{\circ}$, and the symmetry axis of both the disk and the binary are supposed to coincide. The high-resolution
optical spectroscopy of higher Balmer lines indicates gas clouds associated with the stars well within the inner radius of the circumbinary disk \citep{2004A&A...421.1159S}.

V4046~Sgr was observed with the \emph{Chandra} HETGS in two blocks, starting on 6~Aug.~2006 for 100~ks (ObsID 5423) and on 9~Aug.~2006 
for 50~ks (ObsID 6265). First-order grating spectra were extracted by applying standard CIAO 3.3 tools, and then 
positive and negative orders were summed up. Individual emission lines in the HEG and MEG spectra were analysed 
with the CORA line-fitting tool \citep{2002AN....323..129N}, assuming modified Lorentzian line profiles with $\beta=2.5$. 


\begin{center}
\begin{figure}[ht]
\resizebox{\hsize}{!}{\includegraphics{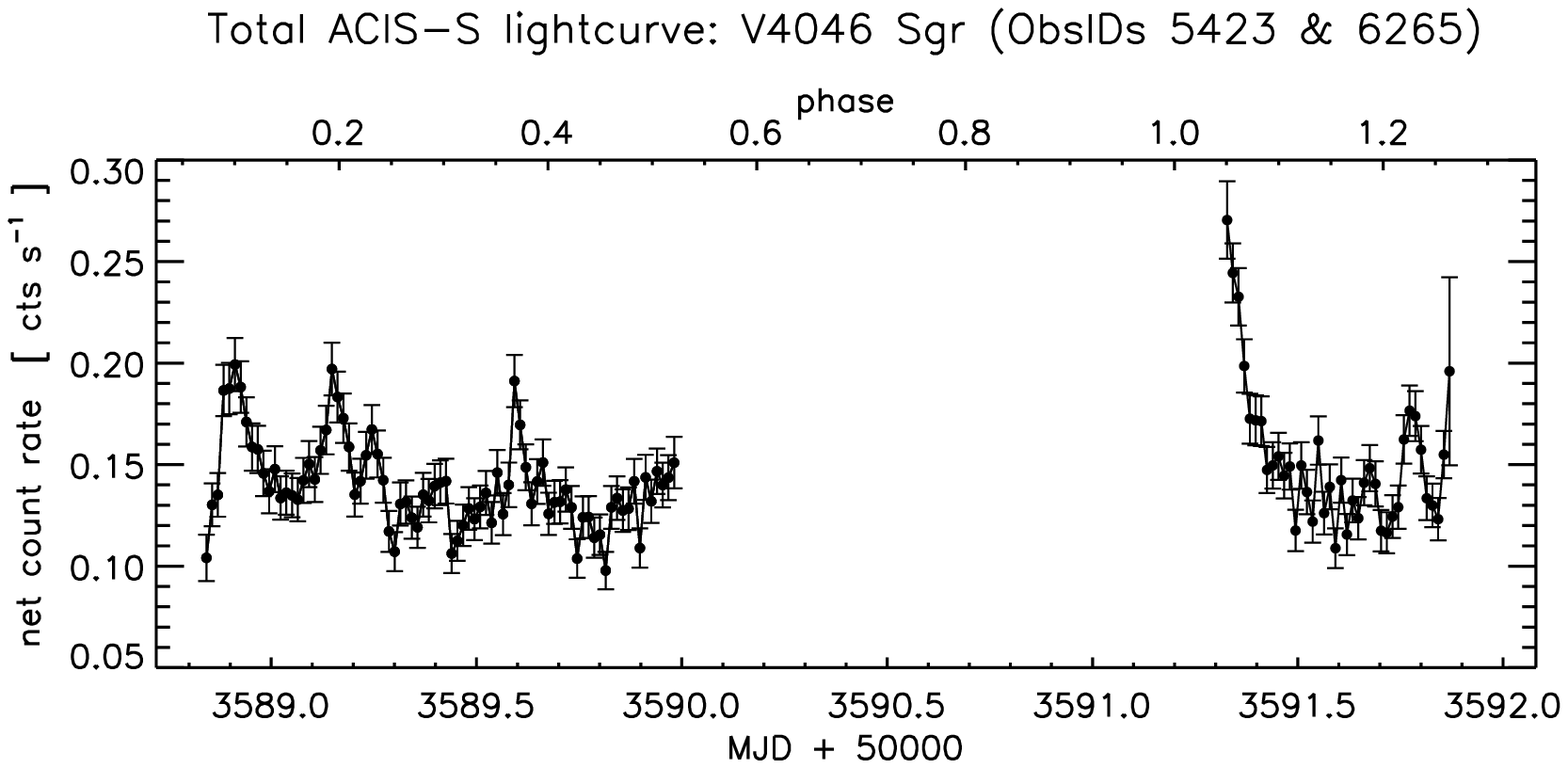}}
\caption{X-ray lightcurve of V4046~Sgr, extracted from the 0$^{\textnormal{\tiny{th}}}$ order data and dispersed spectra. The upper axis shows the orbital phase using the 
ephemerides from \citet{2004A&A...421.1159S}.\label{lightcurve}}
\end{figure}
\end{center}

\section{Results}
The X-ray lightcurve of V4046~Sgr is shown in Fig.~\ref{lightcurve}. Several phases of enhanced flux levels are apparent, where the hardness of the X-ray radiation increases marginally; the second observation starts with what looks like the decay phase of a large flare. Variations in the shape of the higher Balmer line profiles are known to correlate with the orbital 
phase \citep{2004A&A...421.1159S}, but we do not find any obvious correlation between X-ray flux and phase. We analysed single and combined 
spectra and found no significant differences between the two observations; therefore, we use the merged spectra in all of our analysis. 

\begin{figure}[ht]
\resizebox{\hsize}{!}{\includegraphics{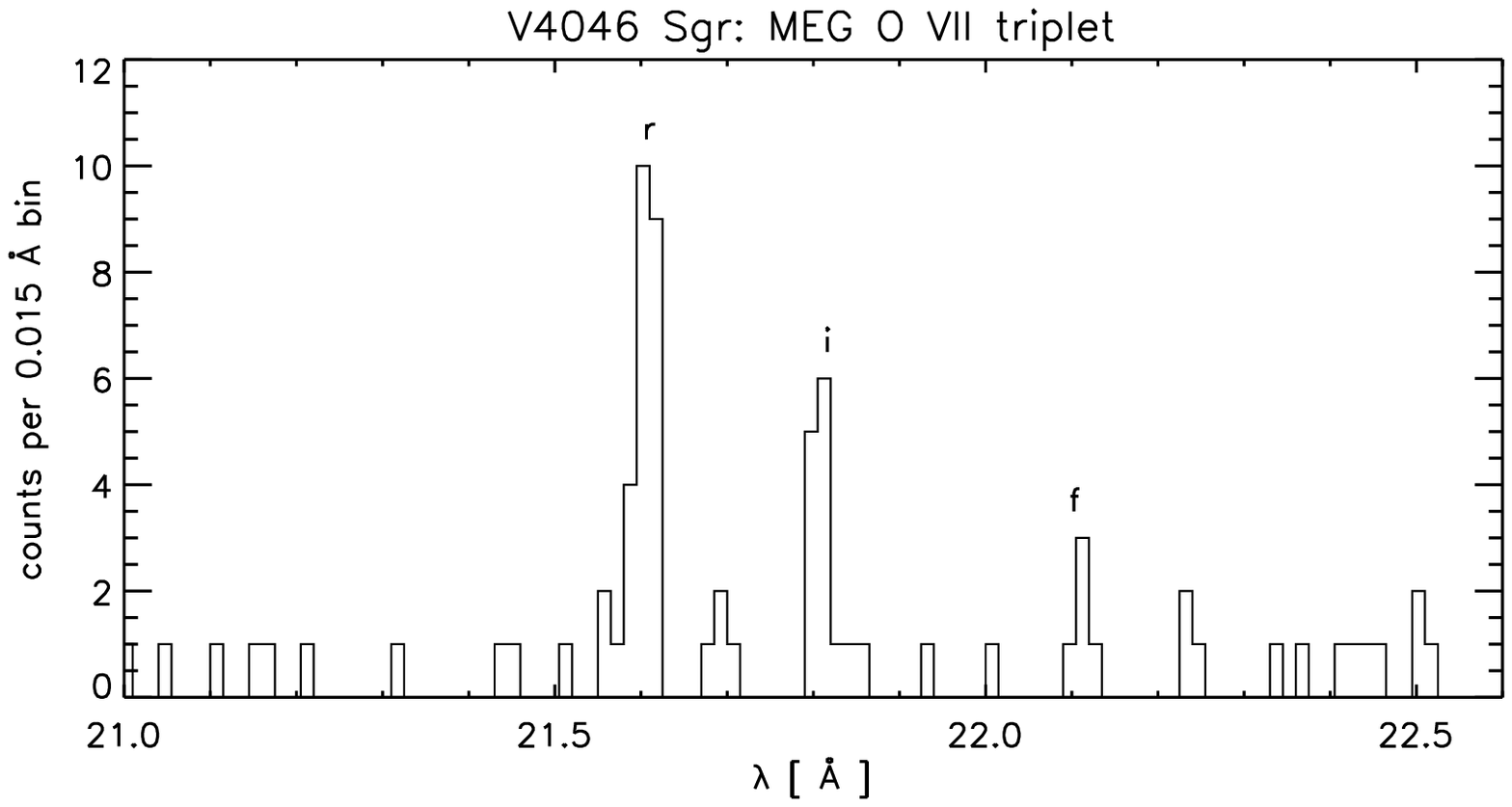}}
\caption{The V4046~Sgr MEG count spectrum around the \ion{O}{vii} triplet region. 
\label{otriplet}}
\end{figure}

The MEG and HEG grating spectra show a typical emission-line spectrum.  Specifically, the Ly\,$\alpha$ lines of 
silicon, magnesium, neon, oxygen, and nitrogen and the Ly\,$\beta$ lines of neon and oxygen are detected (cf. Fig.~\ref{compare}).  Weak
iron lines are found from \ion{Fe}{xvii} and \ion{Fe}{xviii}; He-like resonance lines are detected from oxygen, neon, and silicon, while 
the He-like Mg lines are very weak.  The forbidden and intercombination lines of oxygen, 
neon, and silicon are clearly detected and resolved.
The strongest detected lines, together with the derived best-fit line counts and their errors (as determined with CORA), are listed in Table~\ref{tab1}.

\begin{figure}[ht]
\resizebox{\hsize}{!}{\includegraphics{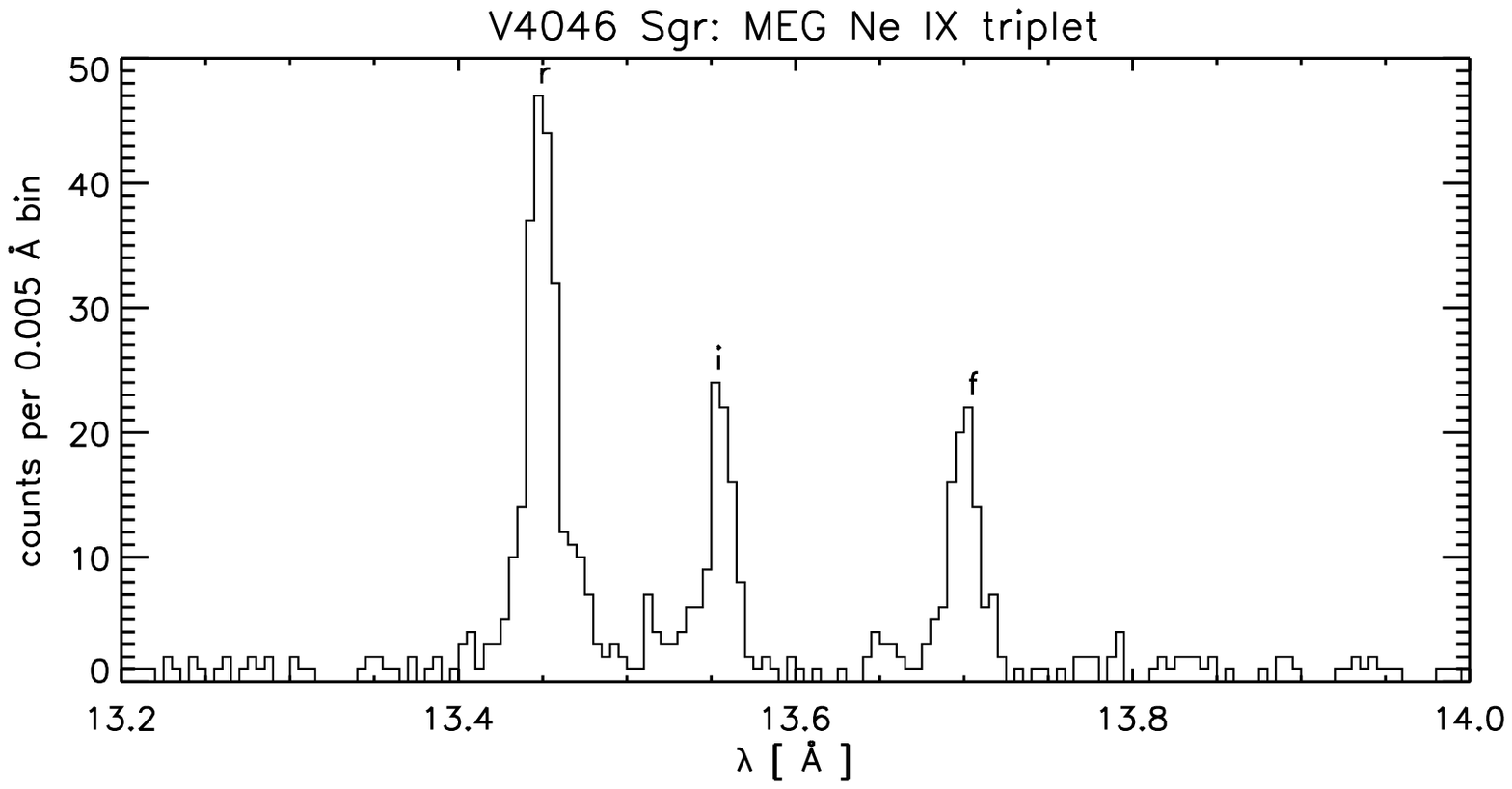}}
\caption{The V4046~Sgr MEG count spectrum around the \ion{Ne}{ix} triplet region. 
\label{netriplet}}
\end{figure}

\begin{figure}
\resizebox{\hsize}{!}{\includegraphics[scale=0.5]{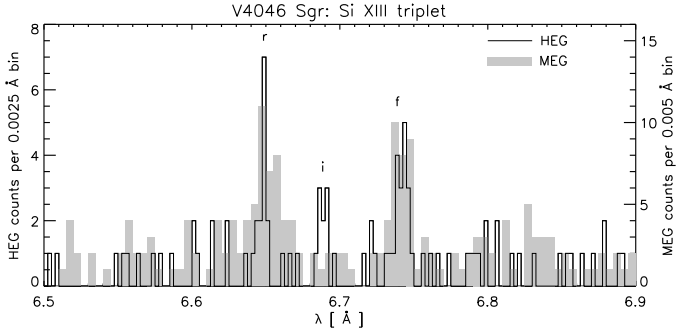}}
\caption{\ion{Si}{xiii} triplet region of V4046~Sgr in the HEG (black line) and MEG (shaded). The intercombination line is clearly stronger in the HEG than in the MEG, but the fits in Table~\ref{tab1} show that both measurements are statistically consistent. \label{sitriplet}}
\end{figure}

In Fig.~\ref{otriplet} we plot the observed MEG count spectrum in the \ion{O}{vii}~triplet region around 22~\AA. 
Obviously, the signal is low, and yet the r and i lines are clearly detected with extremely high significance 
(cf. Table~\ref{tab1}).  However, only $3.0\pm2.2$ counts
are recorded from the f line, clearly indicating deviations from the low-density limit.  We specifically find 
f/i$=0.33\pm0.28$ for \ion{O}{vii}, which must be compared to the low-density limit 
of 3.9 \citep{APECAPED}.
Figure~\ref{netriplet} shows the observed MEG count spectrum in the \ion{Ne}{ix}~triplet region 
around 13.5~\AA. All triplet lines are clearly detected and, as shown by an inspection of the MEG spectrum, 
contamination by \ion{Fe}{xix} \citep[cf.][]{2003ApJ...598.1277N} is of the order of a few percent at its worst
for V4046~Sgr similar to the case of TW~Hya \citep{twhya}.
The \ion{Ne}{ix} f/i-ratio is $1.04\pm0.15$, derived under the assumption of negligible iron contamination, with the \ion{Ne}{ix} low-density limit being 3.1.
Finally, in  Fig.~\ref{sitriplet} we plot the observed HEG and MEG count spectrum in the \ion{Si}{xiii}~triplet 
region around 6.7~\AA.\\
All triplet lines are clearly detected in the HEG spectrum, the f line being approximately 
at the same strength as the r line.  While
only an upper limit for the i line can be obtained from the MEG spectrum, a \ion{Si}{xiii} f/i-ratio
of $1.9\pm0.6$ can be derived from the HEG spectrum, which compares well with the 
\ion{Si}{xiii} f/i-ratios of the larger sample of active stars studied by \citet{2004ApJ...617..508T}.  Since the \ion{Si}{xiii}~f line is located exactly on the Si~K edge of the detector, we investigated whether the observed \ion{Si}{xiii} f/i-ratio (in recorded counts)
is consistent with the \ion{Si}{xiii} f/i-count ratio for a larger sample of stars.   Assuming then that all 
active stars are in the low-density limit for \ion{Si}{xiii} with $\log n_e < 13.5$,
we conclude that the same applies for V4046~Sgr (as far as \ion{Si}{xiii} is concerned).
This conclusion is supported by the ratio of
\ion{Fe}{xvii} 17.10~\AA\,/\,17.05~\AA{} of about unity, which \citet{Ness0510} argue to be the 
low-density limit ($\log n_e<13.6$); however, uncertain atomic physics require some caution.

\begin{table} [h]
\caption{Measured line fluxes for V4046~Sgr \label{tab1}}
\begin{center}
\begin{tabular}{lrrrrr}
\hline \hline
Line ID                 & $\lambda$     & MEG           & HEG           & Photon         \\
                        & [ \AA\ ]         & [ counts ]    &[ counts ]     & flux $^a$ \\
\hline
\ion{Si}{xiv} Ly\,$\alpha$& 6.18        & $32  \pm 6  $ &$15  \pm 4    $&$2.5\pm0.7 $           \\
\ion{Si}{xiii} r        & 6.64          & $46 \pm 8  $  &$15  \pm 4    $&$2.7\pm0.5 $           \\
\ion{Si}{xiii} i        & 6.69          & $10^b$        &$9   \pm 3    $&$1.5\pm0.5 $           \\
\ion{Si}{xiii} f        & 6.74          & $36 \pm 7  $  &$20   \pm 5   $&$2.9\pm0.7 $           \\
\ion{Mg}{xii} Ly\,$\alpha$& 8.42        & $29  \pm 6  $ & n.a.          &$1.6\pm0.4 $           \\
\ion{Mg}{xi} r          & 9.17          & $35 \pm 7  $  & n.a.          &$2.7\pm0.5 $           \\
\ion{Ne}{x} Ly\,$\beta$   &10.23         & $ 57\pm8   $  & n.a.          &$ 5.2\pm0.7 $          \\
\ion{Ne}{x} Ly\,$\alpha$  &12.14         & $280\pm17  $  &$102\pm10$     &$48.7\pm3.0 $          \\
\ion{Ne}{ix} r          &13.46          & $245 \pm16  $ &$64 \pm 8  $   &$74.2\pm4.8 $          \\
\ion{Ne}{ix} i          &13.56          & $107\pm 11 $  &$26\pm 5    $  &$32.0\pm3.3 $          \\
\ion{Ne}{ix} f          &13.70          & $105\pm 11 $  &$23 \pm 5    $ &$33.2\pm3.3 $          \\
\ion{O}{viii} Ly\,$\beta$ &16.02          & $23 \pm  5 $  & n.a.          &$ 11.0  \pm   2.0 $    \\
\ion{O}{viii} Ly\,$\alpha$&18.97          & $98 \pm 10 $  & n.a.          &$115.7\pm11.8 $        \\
\ion{O}{vii} r          &21.6           & $21 \pm 5  $  & n.a.          &$49.3\pm11.7 $         \\
\ion{O}{vii} i          &21.8           & $10.4\pm 3.5$ & n.a.          &$26.8\pm9.0 $          \\
\ion{O}{vii} f          &22.1           & $3.0 \pm 2.2$ & n.a.          &$9.0\pm7.0 $           \\
\ion{N}{vii} Ly\,$\alpha$ &24.78          & $28 \pm  6 $  & n.a.          &$ 75.0\pm16.1 $        \\
\ion{Fe}{xvii}          &15.01          & $38 \pm 7  $  & n.a.          &$17.0\pm3.2 $          \\
\ion{Fe}{xvii}          &15.26          & $21 \pm 5 $   & n.a.          &$ 8.1\pm1.9 $          \\
\ion{Fe}{xvii}          &17.05          & $21 \pm 5  $  & n.a.          &$14.0\pm0.3 $          \\
\ion{Fe}{xvii}          &17.10          & $22 \pm 5 $   & n.a.          &$14.6\pm0.3 $          \\
\hline
\end{tabular}
\end{center}
$^a$ in units of $10^{-6}$~photons~cm$^{-2}$~s$^{-1}$ calculated from the MEG data except for the \ion{Si}{xiii} triplet, where only an upper limit is available for the \ion{Si}{xiii}~i line\\
$^b\ 2\sigma$ upper limit
\end{table}

\section{Discussion}
The X-ray data on the CTTS TW~Hya and BP~Tau have been interpreted in terms of an accretion funnel scenario,
where the X-ray emission is emitted in a shock (``hot spot'') produced by the infall of material  along the magnetic 
field essentially at free-fall velocity onto the stellar surface.  The main reason for this interpretation
was the anomalously low \ion{O}{vii} f/i-ratio observed for TW~Hya and BP~Tau, since the
extensive spectral survey by \citet{2004A&A...427..667N} found no star that would even come close to the low  \ion{O}{vii} f/i-ratio observed for 
those two stars. The question is therefore, wether the very same scenario also applies to the CTTS V4046~Sgr and possibly to CTTS as a class?  

\begin{figure}[ht]
\resizebox{\hsize}{!}{\includegraphics[scale=0.5]{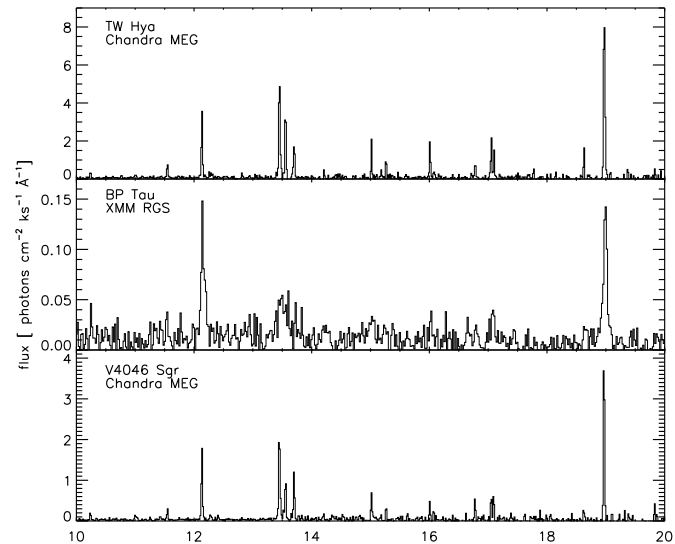}}
\caption{Comparison of the X-ray flux spectra of TW~Hya, BP~Tau, and V4046~Sgr.\label{compare}}
\end{figure}

In Fig.~\ref{compare} we compare the X-ray spectra in the range 10~\AA\ to 20~\AA\ for the three TW~Hya (MEG), 
BP~Tau (\emph{XMM} RGS), and CTTS V4046~Sgr (MEG).  Clearly, the energy resolution of the RGS instrument and the SNR of the BP~Tau data set
is lower than that of the MEG spectra. Nevertheless, some common trends, 
as well as differences, appear:  In all three cases, the \ion{O}{viii} Ly\,${\alpha}$ line
is the strongest line, and neon lines are stronger than iron lines, which are weak if not absent.  In BP~Tau 
the \ion{Ne}{x}  Ly\,$\alpha$ line is much stronger than the \ion{Ne}{ix} He-like r line, 
the opposite applies to TW~Hya and V4046~Sgr.
The different  Ly\,$\alpha$ to He-like resonance line ratios for our three CTTS
are summarised in Table~\ref{lya2r}.  Interpreting these ratios in terms
of an effective temperature, 
 we note that the emission measure in V4046~Sgr cannot 
be described by a single temperature emission model. 
We also investigated the Ly\,$\alpha$\,/\,Ly\,$\beta$ ratios for \ion{Ne}{x} and \ion{O}{viii} and compared the temperatures obtained in this way to estimates from the ratio of Ly\,$\alpha$ to He-like resonance line. The temperatures 
found in this way 
are both about 0.1~dex below the peak formation temperatures of their respective ionisation stages. We find that the ratios are fully consistent 
with optically thin emission, even if the absorption to V4046~Sgr is as large as $N_H=10^{21}$~cm$^{-2}$. A second tracer of 
resonant scattering is the ratio \ion{Fe}{xvii} 15.26~\AA\,/\,15.01~\AA, which matches optically thin predictions as well.

With regard to the He-like f/i-ratios for TW~Hya, BP~Tau, and V4046~Sgr (cf. Table~\ref{lya2r}),  
we find the observed \ion{O}{vii} f/i-ratio of V4046~Sgr to
compare well with what is observed for BP~Tau and to be fully consistent with the lower value of TW~Hya. When interpreted as a pure
density diagnostics (i.e., assuming no contaminating UV flux), we find
$\log n_e\approx 11.5$ \citep{APECAPED}.
Thus TW~Hya, BP~Tau, and V4046~Sgr are in marked contrast to the CTTS T~Tau, where \citet{guedel06} measured 
an f/i ratio of $\sim 4$ and inferred an upper limit to the electron density of $1.4\times 10^{10}$~cm$^{-3}$ (68\% confidence).
\ion{Ne}{ix} is formed at higher temperatures, and the \ion{Ne}{ix} f/i-ratio of V4046~Sgr
is somewhat higher than those measured for TW~Hya and BP~Tau, but still 
smaller than all the 48 \ion{Ne}{ix} f/i-ratios
determined by \citet{2004A&A...427..667N} for their sample of active stars.  The formal \ion{Ne}{ix} density derived for  V4046~Sgr is $\log n_e\approx 12.0$.
We stress that the f/i ratios for both \ion{Ne}{ix} and \ion{O}{vii} 
(peak formation temperatures $\log T\approx6.6$ and $\log T\approx6.3$, respectively) clearly 
deviate from the low-density limit; hence, the X-ray flux is unlikely to be produced in a stellar corona.
On the contrary, the f/i-ratio measured for the \ion{Si}{xiii} triplet agrees with the \ion{Si}{xiii} f/i-ratios 
found in the survey by \citet{2004ApJ...617..508T}; the \ion{Si}{xiii} peak formation temperature of  $\log T\approx 7.0$ 
cannot be reached in an accretion shock, so this triplet probably has a coronal origin.

Turning now to elemental abundances we use ratios of linear combinations of the measured fluxes of the H-like Ly$\alpha$ and He-like resonance lines to derive abundance ratios that are, as a first approximation, independent of the underlying temperature structure of the emitting plasma \citep{alphacen}.  In  Table~\ref{lya2r} we list the values of $A_{\rm{Si}}$/$A_{\rm{Mg}}$, $A_{\rm{Mg}}$/$A_{\rm{Ne}}$, and $A_{\rm{Ne}}$/$A_{\rm{O}}$ obtained for the three CTTS, as well as the solar value \citep{1998SSRv...85..161G} for comparison.
 \citet{2005ApJ...627L.149D} found the Ne/O abundance in BP~Tau consistent with their sample of post-T~Tauri stars, while \citet{twhya} and \citet{bptau} interpret the -- compared to solar -- enhanced Ne/O abundances as due to grain depletion.  Also,
V4046~Sgr shows the same enhanced Ne/O abundances as found for TW~Hya and BP~Tau.
As a consequence, the Mg/Ne abundance is low, both for V4046~Sgr and for TW~Hya; however, the relative Si/Mg abundances are comparable to solar values. 

\begin{table} [ht]
\caption{\label{lya2r} Comparison of properties of the three CTTS TW~Hya, BP~Tau, and V4046~Sgr}
\begin{center}
\begin{tabular}{lcccc}
\hline \hline
                & TW~Hya        & BP~Tau        & V4046~Sgr     &\\
\hline
&\multicolumn{3}{c}{f/i ratios}& $\log n_e$\\
\hline
\ion{Si}{xiii}  & n.a           & n.a.          & $1.9\pm0.6$   & $<13.5$\\
\ion{Ne}{ix}    &$0.33\pm0.24$  &$0.40\pm0.26$  & $1.05\pm0.15$ & $12.0$\\
\ion{O}{vii}    &$0.05\pm0.05$  &$0.37\pm0.16$  &$0.33\pm0.28$  & $11.5$\\
\hline
&\multicolumn{3}{c}{Ly\,$\alpha$\,/\,r ratios}& $\log T$ [ K ] \\
\hline
Si      &$1.83\pm0.61$& n.a.  & $1.13\pm0.27$   &$7.10\pm0.05 $\\
Mg      &$0.60\pm0.30$& n.a.  & $1.03\pm0.30$   &$6.91\pm0.05 $\\
Ne      &$0.63\pm0.07$&$2.51\pm0.85$& $0.70\pm0.09$ &$6.61\pm0.01 $\\
O       &$2.02\pm0.41$&$1.59\pm0.35$& $2.36\pm0.60$   &$6.51\pm0.05 $\\
\hline
&\multicolumn{3}{c}{abundance ratios}&solar\\
\hline
$A_{Si}$/$A_{Mg}$&$1.63\pm0.68$&n.a.&$1.58\pm0.36$&0.93\\
$A_{Mg}$/$A_{Ne}$&$0.05\pm0.02$&n.a.&$0.08\pm0.02$&0.32\\
$A_{Ne}$/$A_O$&$0.86\pm0.11$&$0.66\pm0.20$&$1.04\pm0.14$&0.18\\
\hline
\end{tabular}
\end{center} 

\end{table}

\section{Conclusions}

For single CTTS, the model of an accreting hot spot, where material impacts onto the 
stellar surface at free-fall velocity, closely matches  
the observational data of both TW~Hya \citep{2002ApJ...567..434K,twhya} and BP~Tau \citep{bptau}.  Detailed simulations \citep{calvetgullbring,lamzin,acc_model} predict the emissivity, the 
line ratios, the UV field, and the veiling continuum.  V4046~Sgr fits fully in this picture. Optically thin emission from an 
accretion hot spot explains the unusually low f/i-ratios observed 
in \ion{O}{vii} and  \ion{Ne}{ix}. A ``normal'' 
corona accounts for the more 
energetic X-rays and, in particular, for the low-density \ion{Si}{xiii} triplet, as well as for the variability of V4046~Sgr.  This is somewhat
surprising since the accretion geometry of V4046~Sgr is far more complicated because of
its binary nature.
Gas clouds close to the stars can be traced by line profile analysis of higher Balmer lines and seem to exist in the vicinity of the stars,
yet no circumstellar disk should exist due to the small binary separation. 
The variability of the X-ray lightcurve suggests energetic events as observed for the case of BP~Tau, as well
as in the coronae of other stars; and we caution that densities of up to a few $10^{11}$~cm$^{-3}$ \citep{2002ApJ...580L..73G} seem to be occasionally reached in flares. As to abundances, V4046 Sgr shows the very high Ne/O abundance, reminiscent of TW Hya,
and grain depletion is a suggestive scenario to explain the apparently high Ne abundance.

In summary, three of four CTTS with available high-resolution and high SNR X-ray grating spectra show
far lower f/i ratios in the \ion{O}{vii} and \ion{Ne}{ix} triplets than any other star. In  V4046~Sgr we are likely to see the emission
from a corona in the high-temperature Si lines, and the binary nature of V4046~Sgr seems to have little influence on its 
X-ray spectrum. 
We thus conclude that the presence, not the absence, of low f/i ratios appears to be more typical of CTTS.
Also, the emission mechanism obviously works in single and close binary 
stars, presenting a challenge for future magnetohydrodynamic simulations of the accretion funnel.

\begin{acknowledgements}
HMG, CL, and JR acknowledge support from the DLR under grant 50OR0105. J.-U. N. gratefully acknowledges support provided by NASA through Chandra
Postdoctoral Fellowship grant PF5-60039 awarded by the Chandra X-ray 
Center, which is operated by the Smithsonian Astrophysical Observatory for 
NASA under contract NAS8-03060.
\end{acknowledgements}

\bibliographystyle{aa} 
\bibliography{../articles}

\end{document}